\documentclass{PoS}
\usepackage{epsfig}

\PoS{PoS(LAT2005)357}

\title{Kaon semileptonic decay form factors in two-flavor QCD}

\ShortTitle{Kaon semileptonic decay form factors in two-flavor QCD}

\author{
JLQCD Collaboration:\\
\speaker{N.~Tsutsui}${}^a$\thanks{E-mail: naoto.tsutsui@kek.jp},
S.~Aoki${}^b$, M.~Fukugita${}^c$, S.~Hashimoto${}^{a,d}$, K-I.~Ishikawa${}^e$, N.~Ishizuka${}^{b,f}$,
Y.~Iwasaki${}^b$, K.~Kanaya${}^b$, T.~Kaneko${}^{a,d}$, Y.~Kuramashi${}^{b,f}$, M.~Okawa${}^e$,
A.~Ukawa${}^{b,f}$, N.~Yamada${}^{a,d}$, and T.~Yoshi\'e${}^{b,f}$\\
${}^a$High Energy Accelerator Research Organization (KEK), Tsukuba 305-0801, Japan.\\
${}^b$Graduate School of Pure and Applied Sciences, University of Tsukuba, Tsukuba 305-8571, Japan.\\
${}^c$Institute for Cosmic Ray Research, University of Tokyo, Kashiwa 277-8572, Japan.\\
${}^d$School of High Energy Accelerator Science, The Graduate University for Advanced Studies (Sokendai), Tsukuba 305-0801, Japan.\\
${}^e$Department of Physics, Hiroshima University, Higashi-Hiroshima 739-8526, Japan.\\
${}^f$Center for Computational Sciences, University of Tsukuba, 305-8577, Japan.\\
}

\abstract{
We present a calculation of the kaon form factors in two-flavor QCD with
the non-perturbatively $O(a)$-improved Wilson quark action. In order to
achieve a few percent accuracy in the study of SU(3) breaking effects,
we use a set of double ratios of the matrix elements, with which the
bulk of the statistical fluctuation and the multiplicative
renormalization factors cancel.
}

\FullConference{XXIIIrd International Symposium on Lattice Field Theory\\
                25-30 July 2005\\
                Trinity College, Dublin, Ireland}

\begin{document}
\section{Introduction}
One of the most important quantities in the low energy kaon physics is
the form factor $f_+$ at zero momentum transfer, which provides a
theoretical input for the determination of the Kobayashi-Maskawa matrix
element $|V_{us}|$ through the $K_{l3}$ decay.

For this form factor it is known that
\begin{enumerate}
 \item The leading contribution is unity, which is exact in the SU(3) limit,
 \item There is no contribution of $O(m_s-\bar{m})$ due to the
       Ademollo-Gatto theorem where $\bar{m}=(m_u+m_d)/2$.
\end{enumerate}
In 1984, Leutwyler and Roos estimated this form factor analytically
\cite{Leutwyler:1984je} and their result $f_+(0)=0.961\pm0.008$ has been
used as the standard in the phenomenological analysis. Though the
leading correction was determined unambiguously using chiral
perturbation theory, the next-to-leading order correction was estimated
using a model of the wave function of the pseudoscalar meson, with which
only a crude error estimate is available. The numerical simulation of
lattice QCD could help this situation. Indeed, a quenched result appears
recently~\cite{Becirevic:2004ya} and several groups carry out the
unquenched calculation with different lattice setups~\cite{other}.

Our calculation is performed on two-flavor QCD gauge configurations
generated with the non-perturbatively $O(a)$-improved Wilson fermion by
the JLQCD Collaboration~\cite{Aoki:2002uc}. The lattice size is $20^3
\times 48$ at $\beta=5.2$ and 12,000 HMC trajectories are
accumulated. Five sea quark masses corresponding to $m_{PS}/m_V \sim
0.8-0.6$ enable us to make a careful study of the quark mass dependence
of the form factors. Especially, we focus on the consistency between the
lattice data and the ChPT predictions. The pion form factors are
obtained as a by-product of the calculation. The results are presented
in the poster session by S.~Hashimoto~\cite{Hashimoto:lattice2005}.

\section{Kaon form factors}
The kaon semileptonic decay form factors $f_+(q^2)$ and $f_-(q^2)$ are
defined as
\begin{equation}
\langle \pi(\vec{p}) | V_\mu | K(\vec{k}) \rangle =
f_+(q^2)(k+p)_\mu + f_-(q^2)(k-p)_\mu,
\end{equation}
where $q^2=(k-p)^2$ and $V_\mu$ is the vector part of the weak
current. Reliable theoretical estimate of $f_+(0)$ is necessary for the
precise determination of the Kobayashi-Maskawa matrix element $|V_{us}|$
from the $K_{l3}$ decay. Our goal is to calculate the form factor in a
few percent accuracy from the first principle of QCD. To do that,
careful study of SU(3) breaking effects and the chiral extrapolation are
vital.

The scalar form factor $f_0(q^2)$ can be expressed as a linear
combination of $f_+(q^2)$ and $f_-(q^2)$
\begin{equation}
f_0(q^2) = f_+(q^2) + \frac{q^2}{m_K^2-m_\pi^2}f_-(q^2),
\end{equation}
and the ratio of $f_-(q^2)$ and $f_+(q^2)$ is denoted as $\xi(q^2)$
\begin{equation}
\xi(q^2) = \frac{f_-(q^2)}{f_+(q^2)}.
\end{equation}

\section{Lattice calculation}
Our calculation proceeds in three steps, and the form factor $f_+(0)$ is
expressed by a product of three factors
\begin{equation}
f_+(0) =
f_0(q_{max}^2) \times
\frac{f_+(0)\left[1+\xi(0)\displaystyle\frac{m_K-m_\pi}{m_K+m_\pi}\right]}
     {f_0(q_{max}^2)} \times
\frac{1}{1+\xi(0)\displaystyle\frac{m_K-m_\pi}{m_K+m_\pi}}.
\label{eq:three_steps}
\end{equation}
In the first step, the scalar form factor at the maximum momentum
transfer squared $q_{max}^2=(m_K-m_\pi)^2$ is estimated and the
interpolation to $q^2=0$ is done in the second step. The unnecessary
contribution from $\xi(0)$ is subtracted in the last step. A double
ratio of the three-point functions is used to obtain each factor. We
explain the details of each step in the following subsections.

\subsection{Double ratio I}
\begin{figure}
\centering
\epsfig{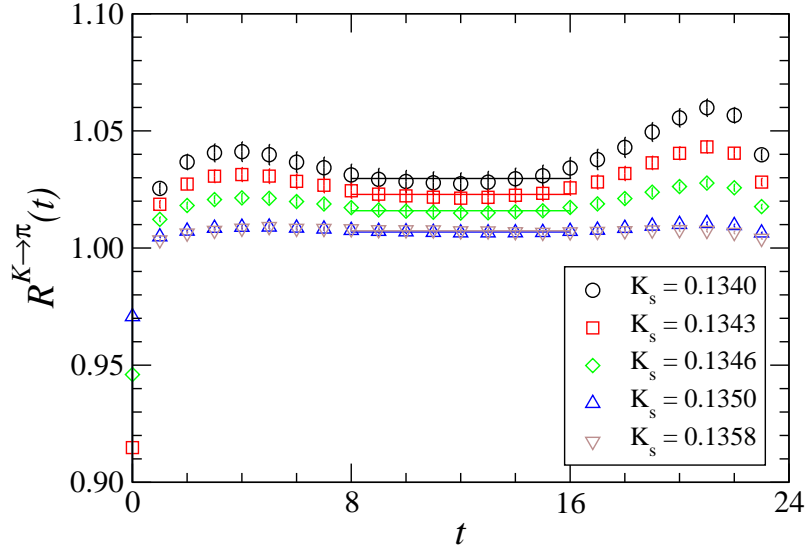}
\caption{Double ratio $R^{K\to\pi}(t)$ for the scalar form factor
 $f_0(q_{max}^2)$ at $K_{sea}=0.1355$.}
\label{fig:double_ratio_k1355}
\end{figure}%

The scalar form factor at the maximum momentum transfer squared
$f_0(q_{max}^2)$ is obtained from the double ratio of three-point
functions
\begin{equation}
R^{K\to\pi}(t) =
\frac{C_{\pi V_4 K}(t,T/2;\vec{0},\vec{0}) C_{K V_4 \pi}(t,T/2;\vec{0},\vec{0})}
     {C_{\pi V_4 \pi}(t,T/2;\vec{0},\vec{0}) C_{K V_4 K}(t,T/2;\vec{0},\vec{0})}
\rightarrow
\left[
\frac{m_K+m_\pi}{2\sqrt{m_Km_\pi}}f_0(q_{max}^2)
\right]^2,
\;\;\;\;0 \ll t \ll T/2,
\label{eq:double_ratio_I}
\end{equation}
where the three-point function is given by
\begin{equation}
  C_{\pi V_\mu K}(t_x,t_y;\vec{p},\vec{q}) =
 \sum_{\vec{x},\vec{y}}
 \langle
 O_\pi(t_y,\vec{y})V_\mu(t_x,\vec{x})O_K(0)
 \rangle
 e^{+i\vec{q}\cdot\vec{x}} e^{-i\vec{p}\cdot\vec{y}},
\end{equation}
and $O_\pi$, $O_K$ is the interpolating operator of pion and kaon. This
double ratio is essentially the same as that used before in the
calculation of the $B \to Dl\nu$ form factor by the Fermilab
group~\cite{Hashimoto:1999yp}.

A typical time dependence of the double ratio $R^{K\to\pi}(t)$ at the
lightest sea quark mass $K_{sea}=0.1355$ is plotted in
Figure~\ref{fig:double_ratio_k1355}. Smeared kaon and pion sources are
set at time $t=0,T/2$, respectively. The vector current is inserted at
$t$. Because not only the bulk of the statistical fluctuation but the
multiplicative renormalization factors and other systematic errors
cancel, we can determine $f_0(q_{max}^2)$ with an accuracy better than
one percent.

\subsection{Double ratio II}
The second double ratio with the pion two-point function
$C_{\pi\pi}(t;\vec{p})$
\begin{equation}
\frac{\displaystyle\frac{C_{\pi V_4 K}(t,T/2;\vec{p},\vec{p})}
                        {C_{\pi V_4 K}(t,T/2;\vec{0},\vec{0})}}
     {\displaystyle\frac{C_{\pi\pi}(t;\vec{p})}
                        {C_{\pi\pi}(t;\vec{0})}}
\to
\frac{m_K+E_\pi(\vec{p})}{m_K+m_\pi}
\frac{f_+(q^2)\left[1+\xi(q^2)\displaystyle\frac{m_K-E_\pi(\vec{p})}
                                                {m_K+E_\pi(\vec{p})}\right]}
     {f_+(q_{max}^2)\left[1+\xi(q_{max}^2)\displaystyle\frac{m_K-m_\pi}
                                                            {m_K+m_\pi}\right]},
\;\;\;\;0 \ll t \ll T/2,
\label{eq:double_ratio_II}
\end{equation}
is used to interpolate the form factor to $q^2=0$. This is the second
factor in (\ref{eq:three_steps}), because
$f_0(q_{max}^2)=f_+(q_{max}^2)\left[1+\xi(q_{max}^2)
\frac{m_K-m_\pi}{m_K+m_\pi}\right]$.
In this case, the momentum is inserted at $t$ and $T/2$ so that the pion
has a finite momentum and the kaon is at rest. The double ratio
(\ref{eq:double_ratio_II}) is statistically noisier than the double
ratio (\ref{eq:double_ratio_I}) in the previous subsection. The data
corresponding to the pion momenta $\vec{p}=2\pi/20\times(\pm1,0,0)$,
$(\pm1,\pm1,0)$ and $(\pm1,\pm1,\pm1)$ are obtained and the quadratic
function is used to interpolate to $q^2=0$, which is shown in
Figure~\ref{fig:xi_k1355} (left).

\subsection{Double ratio III}
\begin{figure}
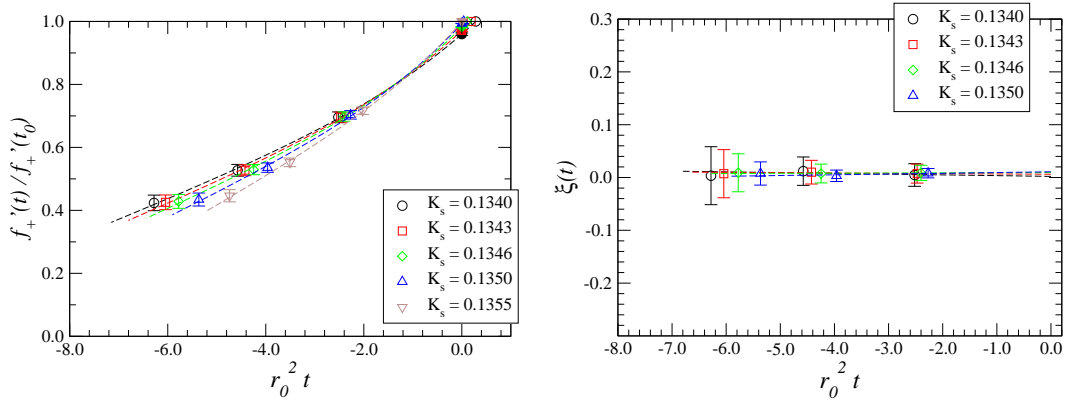

\centering
\begin{tabular}{cc}
\epsfig{file=factor_k1355.eps, width=.45\textwidth,clip} &
\epsfig{file=xi_k1355.eps, width=.45\textwidth,clip}
\end{tabular}
\caption{The ratio of form factor $f_+^\prime(t)/f_+^\prime(t_0)$ (left)
and $\xi(t)$ (right) as a function of momentum transfer squared $t=q^2$ at
$K_{sea}=0.1355$. A notation
$f_+^\prime(t)=f_+(t)\left[1+\xi(t)\frac{m_K-E_\pi(\vec{p})}{m_K+E_\pi(\vec{p})}\right]$
is used here.}
\label{fig:xi_k1355}
\end{figure}%

The last double ratio is for the determination of $\xi(0)$
\begin{equation}
\frac{C_{\pi V_i K}(t,T/2;\vec{p},\vec{p}) C_{\pi V_4 \pi}(t,T/2;\vec{p},\vec{p})}
     {C_{\pi V_4 K}(t,T/2;\vec{p},\vec{p}) C_{\pi V_i \pi}(t,T/2;\vec{p},\vec{p})}
\to
\frac{1-\xi(q^2)}{\displaystyle\frac{m_K+E_\pi(\vec{p})}{m_\pi+E_\pi(\vec{p})} +
\xi(q^2)\frac{m_K-E_\pi(\vec{p})}{m_\pi+E_\pi(\vec{p})}},
\;\;\;\;0 \ll t \ll T/2,
\end{equation}
where we have to measure the spatial component of the matrix
element. The extrapolation to $q^2=0$ is done by assuming the linear
dependence on $q^2$, which is shown in Figure~\ref{fig:xi_k1355}
(right). We see that the $q^2$ dependence of the $\xi$ is very weak and
independent of the strange quark mass.

\section{Result for $f_+(0)$}
\begin{figure}
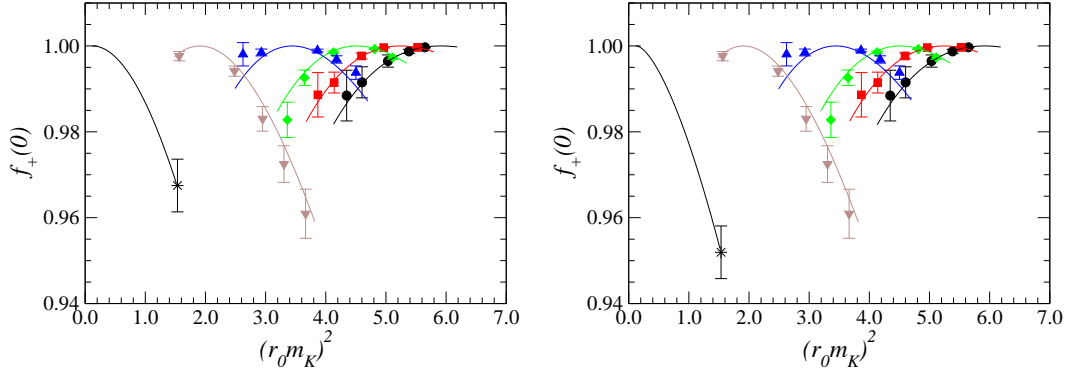

\centering
\begin{tabular}{cc}
\epsfig{file=all_quadfit.eps, width=.45\textwidth,clip} &
\epsfig{file=all_chptfit.eps, width=.45\textwidth,clip}
\end{tabular}
\caption{Chiral extrapolation of the form factor $f_+(0)$ with quadratic
 function (left) and the one-loop ChPT formula plus a quadratic function (right).}
\label{fig:all_quadfit}
\end{figure}%

Following the method explained in the previous subsections, the form
factor $f_+(0)$ is estimated as a function of the up/down and strange
quark masses. In order to obtain the form factor at the physical pion
and kaon masses, we first use a simple fitting function
\begin{equation}
f_+(0) = 1-(c_0+c_1[(r_0m_K)^2+(r_0m_\pi)^2])[(r_0m_K)^2-(r_0m_\pi)^2]^2,
\label{eq:quadfit}
\end{equation}
where the meson masses are normalized by the Sommer scale $r_0$. The
result is shown in Figure~\ref{fig:all_quadfit}. Our data are well
represented by function (\ref{eq:quadfit}) and the preliminary result
$f_+(0)=0.967(6)$ is consistent with Leutwyler-Roos's value
$0.961(8)$~\cite{Leutwyler:1984je}, and the quenched result $0.960(9)$
by Be\'cirevi\'c {\it et al}.~\cite{Becirevic:2004ya}

For small pion and kaon masses the ChPT predicts the mass dependence of
the form factor
\begin{equation}
f_+(0)=1+\frac{3}{2}H_{K\pi}(0)+\frac{3}{2}H_{K\eta}(0),\;\;\;\;
H_{PQ}(0) = -\frac{1}{128\pi^2f^2}
\left[m_P^2 + m_Q^2 + \frac{2m_P^2m_Q^2}{m_P^2-m_Q^2}\ln\frac{m_Q^2}{m_P^2}\right].
\end{equation}
We try to fit the data with
\begin{equation}
f_+(0)=1+\frac{3}{2}H_{K\pi}(0)+\frac{3}{2}H_{K\eta}(0)-
(c_0+c_1[(r_0m_K)^2+(r_0m_\pi)^2])[(r_0m_K)^2-(r_0m_\pi)^2]^2.
\end{equation}
At this order the one-loop ChPT formula has no tunable parameters. The
chiral logarithm is significant only in the region where $m_\pi^2 \ll
m_K^2$, while the data in the region $1/2 \leq m_\pi^2/m_K^2 \leq 2$ are
well approximated by the quadratic form (\ref{eq:quadfit}).
The extrapolated value at
physical pion and kaon mass is affected by the chiral logarithm. The
result is $f_+(0)=0.952(6)$.

\section{Charge radius}
The charge radius $\langle r^2 \rangle$ parametrizes the slope of the
form factor near $q^2=0$
\begin{equation}
f(q^2) = f(0)\left[1 + \frac{1}{6}\langle r^2 \rangle q^2 + \cdots\right].
\end{equation}
The parameter $\lambda=\langle r^2 \rangle m_\pi^2/6$ is often used in
the literature.

\begin{figure}
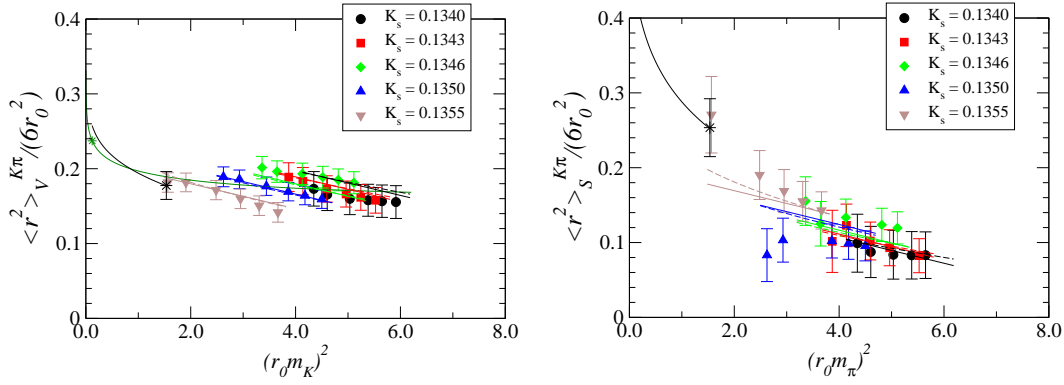

\centering
\begin{tabular}{cc}
\epsfig{file=radius_Kpi.eps, width=.45\textwidth,clip} &
\epsfig{file=scalar_radius.eps, width=.45\textwidth,clip}
\end{tabular}
\caption{Chiral extrapolation of the vector (left) and scalar
 (right) charge radius.}
\label{fig:charge_radius}
\end{figure}

For the vector form factor, our result extrapolated with the one-loop
ChPT plus a quadratic function shown in Figure~\ref{fig:charge_radius}
is $\langle r^2 \rangle^{K\pi}_V=0.26(3)$ fm$^2$. It corresponds to
$\lambda_+=0.021(2)$, which is significantly smaller than the
experimental value 0.0278(7) or the quenched lattice result 0.026(2) by
Be\'cirevi\'c {\it et al}.~\cite{Becirevic:2004ya} Our result for the
charge radius of the scalar form factor (Figure~\ref{fig:charge_radius})
is $\langle r^2 \rangle^{K\pi}_S=0.37(6)$ fm$^2$, which corresponds to
$\lambda_0=0.031(5)$ that overshoots the experimental value 0.0174(22)
or the quenched lattice result 0.012(2).

Our calculation shows that the kaon form factors can be calculated
on the lattice with good precision using the double ratio
method. Further study on systematic errors, especially scaling violation
effects, is needed to obtain the definitive value.

This work is supported by the Large Scale Simulation Program No. 132
(FY2005) of High Energy Accelerator Research Organization (KEK), and
also in part by the Grant-in-Aid of the Ministry of Education
(Nos. 13135204, 13640260, 14046202, 14740173, 15204015, 15540251,
15540279, 15740134, 16028201, 16540228, 17340066).
N.T. is supported by the JSPS Research Fellowship.

\begin{thebibliography}{99}

\bibitem{Leutwyler:1984je}
  H.~Leutwyler and M.~Roos, \emph{Determination of the elements $V_{us}$
	  and $V_{ud}$ of the Kobayashi-Maskawa matrix},
  Z.\ Phys.\ C {\bf 25} (1984) 91.

\bibitem{Becirevic:2004ya}
  D.~Be\'cirevi\'c {\it et al.},
  \emph{The $K\to\pi$ vector form factor at zero momentum transfer on the lattice},
  Nucl.\ Phys.\ B {\bf 705} (2005) 339 [{\tt hep-ph/0403217}].

\bibitem{other}
  C.~Dawson, in these proceedings;
  T.~Kaneko {\it et al.} [RBC Collaboration], in these proceedings;
  M.~Okamoto {\it et al.} [Fermilab, MILC and HPQCD Collaboration],
	[{\tt arXiv:hep-lat/0412044}].

\bibitem{Aoki:2002uc}
  S.~Aoki {\it et al.}  [JLQCD Collaboration],
  \emph{Light hadron spectroscopy with two flavors of O(a)-improved dynamical quarks},
  Phys.\ Rev.\ D {\bf 68} (2003) 054502 [{\tt arXiv:hep-lat/0212039}].

\bibitem{Hashimoto:lattice2005}
  S.~Hashimoto {\it et al.} [JLQCD Collaboration], in these proceedings.

\bibitem{Hashimoto:1999yp}
  S.~Hashimoto, A.~X.~El-Khadra, A.~S.~Kronfeld, P.~B.~Mackenzie, S.~M.~Ryan and J.~N.~Simone,
  \emph{Lattice QCD calculation of $\bar{B} \to D l \bar{\nu}$ decay
	form factors at zero recoil},
  Phys.\ Rev.\ D {\bf 61} (2000) 014502 [{\tt arXiv:hep-ph/9906376}].

\end{thebibliography}
\end{document}